\documentclass[11pt]{article}
\usepackage{amssymb}

   \csname @addtoreset\endcsname{equation}{section}
    \csname @addtoreset\endcsname{figure}{section}
    \csname @addtoreset\endcsname{table}{section}

\newcounter{thanksnum}
\def\thanksnumber#1
{\setcounter{thanksnum}{\value{footnote}}\setcounter{footnote}{#1}%
                     \addtocounter{footnote}{-1}\footnotemark
                     \setcounter{footnote}{\value{thanksnum}}}
\def\newtheoremz#1{\@ifnextchar[{\@othmz{#1}}{\@nthmz{#1}}}

\def\@nthmz#1#2{%
\@ifnextchar[{\@xnthmz{#1}{#2}}{\@ynthmz{#1}{#2}}}

\def\@xnthmz#1#2[#3]{\expandafter\@ifdefinable\csname #1\endcsname
{\@definecounter{#1}\@addtoreset{#1}{#3}%
\expandafter\xdef\csname the#1\endcsname{\expandafter\noexpand
  \csname the#3\endcsname \@thmcountersepz \@thmcounterz{#1}}%
\global\@namedef{#1}{\@thmz{#1}{#2}}\global\@namedef{end#1}{\@endtheoremz}}}

\def\@ynthmz#1#2{\expandafter\@ifdefinable\csname #1\endcsname
{\@definecounter{#1}%
\expandafter\xdef\csname the#1\endcsname{\@thmcounterz{#1}}%
\global\@namedef{#1}{\@thm{#1}{#2}}\global\@namedef{end#1}{\@endtheoremz}}}

\def\@othmz#1[#2]#3{\expandafter\@ifdefinable\csname #1\endcsname
  {\global\@namedef{the#1}{\@nameuse{the#2}}%
\global\@namedef{#1}{\@thmz{#2}{#3}}%
\global\@namedef{end#1}{\@endtheoremz}}}

\def\@thmz#1#2{\refstepcounter
    {#1}\@ifnextchar[{\@ythmz{#1}{#2}}{\@xthmz{#1}{#2}}}

\def\@xthmz#1#2{\@begintheoremz{#2}{\csname the#1\endcsname}\ignorespaces}
\def\@ythmz#1#2[#3]{\@opargbegintheoremz{#2}{\csname
       the#1\endcsname}{#3}\ignorespaces}

\def\@thmcounterz#1{\noexpand\arabic{#1}}
\def\@thmcountersepz{.}
\def\@begintheoremz#1#2{ \trivlist \item[\hskip \labelsep{\bf #1\ #2}]}
\def\@opargbegintheoremz#1#2#3{ \trivlist
      \item[\hskip \labelsep{\bf #1\ #2\ (#3)}]}
\def\@endtheoremz{\endtrivlist}

\newtheorem{theorem}{Theorem}[section]
\newtheorem{lemma}{Lemma}[section]

\newtheorem{corollary}{Corollary}[section]

\def\e{\varepsilon}

\def\defi{\stackrel{{\scriptscriptstyle \Delta}}{=}}

\def\d{\delta}
\def\o{\omega}
\def\O{\Omega}

\def\F{{\cal F}}
\def\w{\widehat}

\def\R{{\bf R}}
\def\E{{\bf E}}
\def\P{{\bf P}}

\def\S{{\bf S}}

\def\J{{\cal J}}

\def\s{\delta}

\def\ww{\widetilde}

\def\t{\theta}
\def\oo{\bar}
\def\s{\sigma}
\newcommand{\be}{\begin{equation}}
\newcommand{\ee}{\end{equation}}
\newcommand{\bd}{\begin{displaymath}}
\newcommand{\ed}{\end{displaymath}}
\newcommand{\ba}{\begin{array}{ll}}
\newcommand{\ea}{\end{array}}
\newcommand{\baa}{\begin{eqnarray}}
\newcommand{\eaa}{\end{eqnarray}}
\newcommand{\baaa}{\begin{eqnarray*}}
\newcommand{\eaaa}{\end{eqnarray*}}
\font\sm=cmr10
\def\oo{\bar}

  \title{
Mutual Fund Theorem for continuous time markets with random
coefficients}
\author{
Nikolai Dokuchaev\\
 {\sm   Department of Mathematics, Trent University, Ontario,
Canada}}
\begin{document}
\maketitle
\begin{abstract} We study the optimal investment problem for a
continuous time incomplete market model such that the risk-free
rate, the appreciation rates and the volatility  of the stocks are
all random; they are assumed to be independent from  the driving
Brownian motion, and  they  are supposed to be currently observable.
It is shown  that some weakened version of Mutual Fund Theorem holds
for this market for general class of utilities; more precisely, it
is shown that the supremum of expected utilities can be achieved on
a sequence of strategies with a certain distribution of risky assets
that does not depend on risk preferences described by different
utilities.
\end{abstract}
%{\it Abbreviated head}: {\sm }
\section{Introduction} We study an optimal portfolio selection
problem  in a market model which consists of a risk--free bond or
bank account and a finite number of risky stocks.  The evolution of
stock prices is described by Ito stochastic differential equations
with the vector of the appreciation rates $a(t)$ and the volatility
matrix $\s(t)$, while the bond price is exponentially increasing
with a random risk free rate $r(t)$. A typical optimal portfolio
selection problem is to find an investment strategy that maximizes
 $\E U(\ww X(T))$, where $\E$ denotes the
mathematical expectation, $U(\cdot)$ is an utility function, $X(T)$
represents the wealth at final time $T$, and $\ww
X(T)=\exp\biggl(-\int_0^Tr(s)ds\biggr)X(T)$ is the discounted
wealth. There are many works devoted to different modifications of
this problem (see, e.g., Merton (1969) and review in Hakansson
(1997) and Karatzas and Shreve (1998)).
\par
Dynamic  portfolio selection problems are usually studied in the
framework of stochastic control. To suggest a strategy, one needs to
forecast future market scenarios (or the probability distributions,
or the future distributions of $r(t)$, $a(t)$ and $\s(t)$).
 Unfortunately, the
nature of financial markets is such that the choice of a hypothesis
about the future distributions is  always difficult  to justify. In
fact, it is still an open question if there is any useful
information in the past prices that helps to predict the future.
Respectively, there are serious reservations toward usual tools of
stochastic control such as Dynamic Programming or Stochastic Maximum
Principle that require knowledge of future $r(t)$, $a(t)$ and
$\s(t)$. It is why some special methods were developed for the
financial models to deal with limited predictability.
\par
One of this tools is the so-called Mutual Fund Theorem that says
that if the distribution of the risky assets in the optimal
portfolio does not depend on the investor's risk preferences (or
utility function). This means that all rational investors may
achieve optimality using the same mutual fund plus a saving account.
Clearly, calculation of the optimal portfolio is easier in this
case.
\par If Mutual Fund Theorem holds, then, for a typical model, portfolio stays
on the {\it efficient frontier} even if there are errors in the
forecast, i.e., it is optimal for some other risk preferences. This
reduces
 the impact of forecast errors. This is another reason why it is
 important to know when  Mutual Fund Theorem
 holds.
 \par
 Mutual Fund Theorem was established first for the
 single period mean variance portfolio selection problem, i.e., for the problem with quadratic criterions.
 This
 result was a cornerstone of the modern portfolio theory. In
 particular, the Capital Assets Pricing Model (CAPM) is based on it.
 For the multi-period discrete time setting, some versions of
Mutua Fund Theorem were obtained so far for problems with quadratic
criterions only (Li and Mg (1999), Dokuchaev (2010)). For the
continuous time setting, Mutual Fund Theorem was obtained for
portfolio selection problems with quadratic criterions as well as
for more general utilities. In particular, Merton's optimal
strategies for $U(x)=\d^{-1}x^\d$ and $U(x)=\log(x)$ are such that
Mutual Fund Theorem holds for the case of random coefficients
independent from the driving Brownian motion
 (Karatzas and Shreve (1998)).
It is also known that Mutual Fund Theorem  does not hold for power
utilities in the presence of correlations (see, e.g., Brennan
(1998), Feldman (2007). Khanna and Kulldorff (1999) proved that
Mutual Fund Theorem theorem holds for a general utility function
$U(x)$ for the case of non-random coefficient, and for a setting
with consumption. Lim (2004) and Lim and Zhou (2002) found some
cases Mutual Fund Theorem theorem holds for problems with quadratic
criterions. Dokuchaev and Haussmann (2001)
 found that Mutual Theorem holds if the scalar value
$\int_0^T|\t(t)|^2dt$ is non-random, where $\t(t)$ is the market
price of risk process.
 Schachermayer {\it et al} (2009) found
sufficient conditions for Mutual Fund Theorem expressed via
replicability of the European type claims $F(Z(T))$, where
$F(\cdot)$ is a deterministic function and $Z(t)$ is the discounted
wealth generated by the log-optimal optimal discounted wealth
process. The required replicability has to be achieved by trading of
the log-optimal mutual fund with discounted wealth $Z(t)$.
\par
It can be summarized that  Mutual Fund Theorem was established so
far for the following continuous time optimal portfolio selection
problems:
 \begin{itemize}
\item[(i)] For $U(x)\equiv \log(x)$ for the case of general random coefficients
$(r,a,\s)$;
\item[(ii)] For  $U(x)=\d^{-1}x^\d$, $\d\neq 0$ for the
random coefficients $(r,a,\s)$ being independent from the driving
Brownian motions;
\item[(iii)] For problems with quadratic criterions;
\item[(iv)]  For
general utility  and for non-random coefficients $(r,a,\s)$;
\item[(v)] For
general utility when the integral $\int_0^T|\t(t)|^2dt$ is
non-random;
\item[(vi)] For
general utility when the claims $F(Z(T))$ can be replicated via
trading of a mutual fund with the discounted wealth $Z(t)$, the
deterministic functions $F$.
 \end{itemize}
 In fact, conditions (iv) or (v) are more restrictive than (vi).
 \par
 Extension of Mutual Fund Theorem on problems
 (i)-(vi) was not trivial; it
 required significant efforts and variety of mathematical methods.
\par
In this paper, we present one more case when Mutual Fund Theorem
holds. More precisely, we found that it holds for general utility
when the parameters $r(t)$, $a(t)$ and $\s(t)$ are all random, they
are independent from the driving Brownian motion, and they are
currently observable.  It is an incomplete market; it is a case of
"totally unhedgeable" coefficients, according to terms from Karatzas
and Shreve (1998), Chapter 6. In fact, we found that only a weakened
version of Mutual Fund Theorem holds: the supremum of expected
utilities can be achieved on a sequence of strategies with a certain
distribution of risky assets that does not depend on utility.
\section{Definitions}
We are  given a standard probability space $(\O,\F,\P)$, where
$\O=\{\o\}$ is a set of elementary events, $\F$ is a complete
$\s$-algebra of events, and $\P$ is a probability measure that
describes a prior probability distributions.
\subsubsection*{Market
model} We consider a market model in a generalized Black-Scholes
framework. We assume that the market consists of a risk free asset
or bank account with price $B(t), $ ${t\ge 0}$, and $n$ risky stocks
with prices $S_i(t)$, ${t\ge 0}$, $i=1,2,\ldots,n$, where
$n<+\infty$ is given.\par  We assume that \be \label{d.B}
B(t)=B(0)\exp\Bigl(\int_0^t r(s)ds\Bigr), \ee where $r(t)$ is the
random process of the risk-free interest rate (or the short rate).
We assume that $B(0)=1$. The process $B(t)$ will be used as
numeraire.
\par The prices of the
stocks evolve according to \be \label{d.S}
dS_i(t)=S_{i}(t)\Bigl(a_i(t)dt+\sum_{j=1}^n\s_{ij}(t) dw_j(t)\Bigr),
\quad t>0, \ee where $w(\cdot)=(w_1(\cdot),\ldots,w_n(\cdot))$ is a
standard Wiener process with independent components, $a_i(t)$ are
the appreciation rates, and $\s_{ij}(t)$ are the volatility
coefficients. The initial price $S_i(0)>0$ is a given non-random
constant.
 \par
We  assume  that $r(t)$, $a(t)\defi\{a_i(t)\}_{i=1}^n$, and
$\s(t)\defi\{\s_{ij}(t)\}_{i,j=1}^{n}$ are currently observable
uniformly bounded, measurable random processes  In addition, we
assume that the inverse matrix $\s(t)^{-1}$ is defined and bounded
and $r(t)\ge 0$.
\par
Let  $\F_t$ be the filtration generated by all observable data. In
particular, we assume that the processes $(S(t),r(t),a(t),\s(t))$ is
adapted to  $\F_t$, where $S(t)\defi (S_1(t),\ldots,S_n(t))^\top$.
\par
Set $\mu(t)\defi (r(t),\ww a(t),\s(t))$, where $\ww a(t)\defi
a(t)-r(t){\bf 1}$ and ${\bf 1}\defi(1,1,\ldots,1)^\top\in\R^n$. The
process $\mu$ represents the vector of current market parameters.
\par
We assume that the process $\mu(t)$ is independent from $w(\cdot)$.
\par Let
$$\ww S(t)=(\ww S_1(t),\ldots,\ww S_n(t))^\top\defi
\exp\left(-\int_0^tr(s)ds\right)S(t). $$ \subsubsection*{Wealth and
strategies} Let $X_0>0$ be the initial wealth at time $t=0$, and let
$X(t)$ be the wealth at time $t>0$, $X(0)=X_0$. Let the process
$\pi_0(t)$ represents the proportion of the wealth invested in the
bond, $\pi_i(t)$ is the proportion of the wealth invested in the
$i$th stock. In other words, the process $\pi_0(t)X(t)$ represents
the proportion of the wealth invested in the bond, $\pi_i(t)X(t)$ is
the proportion of the wealth invested in the $i$th stock,
$\pi(t)=\left(\pi_1(t),\ldots ,\pi_{n}(t)\right)^\top$, $t\ge 0$. We
assume that \be \label{XOBS} \pi_0(t)+\sum_{i=1}^n\pi_i(t)=1, \ee
 The case of negative $\pi_i$ is not excluded.
\par
The process $\ww X(t)\defi \exp\left(-\int_0^tr(s)ds\right) X(t)$
is called the discounted  wealth.
\par
Let ${\bf S}(t)\defi{\rm diag\,} ( S_{1}(t),\ldots, S_{n}(t))$ and
$\ww{\bf S(t)}\defi{\rm diag\,} (\ww S_{1}(t),\ldots,\ww S_{n}(t))$
be the diagonal matrices with the corresponding diagonal elements.
\par
The portfolio is said to be self-financing, if \be \label{in.self1}
dX(t)=X(t)(\pi(t)^\top\S(t)^{-1} dS(t)+\pi_0(t)B(t)^{-1}dB(t)). \ee
\par It follows that for such portfolios
\baa\label{nw}  d\ww X(t)= \ww X(t)\pi(t)^\top\ww\S(t)^{-1} d\ww
S(t), \eaa so $\pi$ alone suffices to specify the portfolio.
\par
 Let \be \label{thetamu}
\t(t)\defi\s(t)^{-1}\ww a(t) \ee
 be the {\em risk premium process}.
\par
 Let $\ww\Sigma(t_1,t_2)$ be the class of
all $\F_t$-adapted processes
  $\pi(\cdot)=(\pi_1(\cdot),\ldots,\pi_n(\cdot)):[t_1,t_2]\times \O\to\R^n$ such that
   $\sup_{t,\o}|\pi(t,\o)|<+\infty$ and that if $\t(t)=0$ then $\pi(t)=0$.
\par
   We shall consider classes
$\ww\Sigma(t_1,t_2)$ as classes of   admissible strategies. For
these strategies, $X(t)>0$ a.e..
\section{The main result}
 Let
$T>0$ and $X_0>0$ be given. Let $U(\cdot):(0,+\infty)\to\R$ be a
given non-decreasing on $(0,+\infty)$ function.
\par
Let
 \baaa
J(\pi)\defi\E U(X(T,0,X_0,\pi)). \eaaa
\par
We will study the problem \be \label{d.costs0} \mbox{Maximize}\quad
J(\pi) \quad\hbox{over}\quad\pi(\cdot)\in\Sigma(0,T) \label{d.syst0}
\ee
\par
Let $\Sigma_{MFT}(t_1,t_2)$ be the set of all strategies
$\pi\in\Sigma(t_1,t_2)$ such that
$\pi(t)^\top=\nu(t)\t(t)^\top\s(t)^{-1}$, where $\nu(t)$ is an one
 dimensional process adapted to $\F_t$.
 \par
\begin{theorem}\label{MFT} Let the function $U$ has the form
 \baa U(x)=U_0(x)-\sum_{k=1}^NU_k(x)x^{-\d_k}+U_{N+1}(x)\log x,
 \label{U}
\eaa where $N\ge 0$ is an integer, $\d_k\in (0,+\infty)$,
$k=1,...,N$, and where continuous functions $U_k:(0,+\infty)\to\R$
are such that $U_{k}(x)\ge 0$, $k=1,...,N+1$,  \baa
&&\inf_{x>0}U_{0}(x)>-\infty, \qquad\quad\nonumber
\\&&\sup_{x>0}U_k(x)<+\infty,\quad k=1,...,N,\qquad\sup_{x\in(0,1)}U_{N+1}(x)<+\infty. \eaa
 Then Mutual Fund Theorem holds in the following sense:
\baa \sup_{\pi\in \Sigma(0,T)} J(\pi)=\sup_{\pi\in
\Sigma_{MFT}(0,T)} J(\pi). \label{eqMFT}\eaa Moreover, there exits a
constant $C>0$ that depends only on $n$ and $\s(\cdot)$ such that
for any $\pi\in\Sigma(0,T)$  and any $\d>0$ there exists a strategy
$\w\pi\in\Sigma_{MFT}(0,T)$ such that \baa
&&J(\w\pi)\ge J(\pi)-\d,\qquad \label{MFTineq0}\\
&&\sup_{t,\o}|\w\pi(t,\o)|\le C\sup_{t,\o}|\pi(t,\o)|.
\label{MFTineq2}\eaa
\end{theorem}
\par Note the class of admissible $U$ is quite wide, with some restrictions
on the order of singularity for utility at $x=0$ in condition
(\ref{U}).
\section{Proofs}
Note  that (\ref{U}) is not required in Lemmas \ref{lemmaDet} and
\ref{k-optimal}.
\begin{lemma}\label{lemmaDet} Let
$\mu(t)=(r(t),\ww a(t),\s(t)$ be a non-random process and let the
function $U$ be non-decreasing and continuous on $(0,+\infty)$. Then
Mutual Fund Theorem holds in the following sense:
 for any  $\pi\in\Sigma(0,T)$ and any
$\d>0$, there exists a strategy $\w\pi\in\Sigma_{MFT}(0,T)$ such
that (\ref{MFTineq0})-(\ref{MFTineq2}) hold and  $$ \w
\pi(t,\o)^\top =\nu(t,\o)\t(t)^\top\s(t)^{-1},
\quad\hbox{where}\quad
\nu(t,\o)=\frac{|\xi(t,\o)\s(t)^\top|}{|\t(t)|},
$$
if $\t(t)\neq 0$, where $\xi(t,\o)$ is a random $n$-dimensional
$\F_t$-adapted process such that $|\xi(t,\o)|\le
\sup_{t,\o}|\pi(t,\o)|$. The constant  $C>0$ in
 (\ref{MFTineq2}) depends only on $n$ and $\s(\cdot)$.
\end{lemma}
\par
{\it Proof of Lemma \ref{lemmaDet}}. Let $\pi\in \Sigma(0,T)$ and
$\d>0$ be given.
 Let
$C\defi\sup_{t,\o}|\pi(t,\o)|$. By the assumptions about
$\Sigma(0,T)$, we  have that $C<+\infty$. Let $\Sigma_C$ be the set
of all strategies from $\ww\pi\in\Sigma(0,T)$ such that
$\sup_{t,\o}|\ww\pi(t,\o)|\le C$.
\par
Consider the optimal control problem with the controlled process
$Y(t)\defi\log \ww X(t)$ and with admissible strategies from
$\Sigma_C$. By Theorem V.2.5 from Krylov (1980), p.225, we obtain
that there exists a so-called {\it Markov strategy} $\pi_M(t)=
F(Y_M(t),t)\in \Sigma_C$, where $ F:\R\times\R\to\R^n$ is a
measurable function such that the closed equation for
$Y_M(t)\defi\log \ww X(t,0,X_0,\pi_M)$ is a diffusion process and
that $J(\w\pi_M)\ge J(\pi)-\d$.
\par
Further, let us apply  the idea of the proof of Theorem 1 from
Khanna and Kulldorff (1999) adjusted to our case of the model
without consumption. Let us select $\w\pi(t)=\w F(Y_M(t),t)\in
\Sigma_{MFT}(0,T)$ such that $\w\pi(t)=f(Y_M,t)$, where the function
$f(x,t):\R^2\to\R$ is defined as a solution of the finite
dimensional maximization problem \baaa \hbox{Maximize}\quad
 f^\top\ww
a(t)\quad \hbox{over}\quad \{f\in\R^n:\quad
|f^\top\s(t)|=|F_M(x,t)^\top\s(t)|\}. \eaaa If $\t(t)\neq 0$ than
$\t(t)\s(t)^{-1}$, then the solution $f=f(x,t)$ is \baa
f^\top=f(x,t)^\top =\t(t)^{\top}\s(t)^{-1}\nu(x,t), \quad
\hbox{where}\quad \nu(x,t)\defi
\frac{|F_M(x,t)^\top\s(t)|}{|\t(t)|}.
 \label{fopt}\eaa
If $\t(t)= 0$ then, by the choice of $\Sigma(0,T)$, we have that
$|F_M(x,t)=0$, and the optimal vector is $f(x,t)=0$.
\par
 We have that \baaa &&\ww X(t,0,X_0,\pi_M)=X_0+\int_{0}^{t}\ww
X(t,0,X_0,\pi_M)\pi_M(s)^\top\ww{\bf S}(s)^{-1}d\ww S(s)
\\&&= X_0\exp\left(\int_{0}^{t}
\left(\pi_M(s)^\top\ww
a(s)-\frac{1}{2}|\pi_M(s)^\top\s(s)|^2\right)ds+\pi_M(s)^\top\s(s)
dw(s)\right). \eaaa Hence  \baaa Y_M(t)=\log
X_0+\int_{0}^{t}\left(F_M(Y_M(s),s)^\top\ww
a(s)-\frac{1}{2}|F_M(Y_M(s),s)^\top\s(s)|^2\right)ds\\+\int_0^tF_M(Y_M(s),s)^\top\s(s)
dw(s). \eaaa
 Let $\w Y(t)\defi \log\ww X(t,0,X_0,\w\pi)$.
We have \baaa \w Y(t)=\log X_0+\int_{0}^{t}
\left(f(Y_M(s),s)^\top\ww a(s)-\frac{1}{2}|f(
Y_M(s),s)^\top\s(s)|^2\right)ds\\+\int_0^tf( Y_M(s),s)^\top\s(s)
dw(s),\eaaa  Let $\xi(t)\defi \w\pi(t)^\top\t(t)-\pi_M(t)^\top\ww
a(t)$. By the choice of $\w\pi$ and $f$, we have that $\xi(t)\ge 0$.
Hence  \baaa \ww Y(t)= \log X_0+\int_{0}^{t}
\left(f(Y_M(s),s)^\top\ww a(s)+\xi(t)-\frac{1}{2}|f(
Y_M(s),s)^\top\s(s)|^2\right)ds\\+\int_0^tf(Y_M(s),s)^\top\s(s)
dw(s). \eaaa It follows that $\w Y(t)=\oo Y(t)+\xi(t)$, where $\oo
Y(t)$ has the same probability distribution as $Y_M(T)$, and
$\xi(t)\ge 0$. It follows that $J(\w\pi)\ge J(\pi_M)\ge J(\pi)-\d$.
\par In addition, we have
$$
\w\pi(t)^\top=|\pi_M(t)^\top\s(t)| e(t)^\top \s(t)^{-1}, \quad
e(t)\defi \frac{\t(t)}{|\t(t)|},\quad \t(t)\neq 0.
$$
Since  $|e(t)|=1$ and the matrix $\s(t)^{-1}$ is bounded, the
estimate (\ref{MFTineq2}) holds. This completes the proof of Lemma
\ref{lemmaDet}. $\Box$.
\par
 Let us consider now the case when the
parameters are predicable on a some given finite horizon.
\begin{lemma}\label{k-optimal} Let  $U$ be non-decreasing and continuous on $(0,+\infty)$, and let there exists
a finite set $\{t_k\}_{k=0}^N$ such that
 $0=t_0<t_1<...<t_N=T$ and such that the values $\mu(t)|_{t\in[t_k,t_{k+1})}$ can be
 predicted at times $t_k$, meaning that  $\mu(t)$ is $\F_{t_k}$-measurable for
 $t\in[t_k,t_{k+1})$, $k<N$. Then Mutual Fund Theorem
holds in the following sense:
 for any  $\pi\in\Sigma(0,T)$ and any
$\d>0$, there exists a strategy $\w\pi\in\Sigma_{MFT}(0,T)$ such
that (\ref{MFTineq0})-(\ref{MFTineq2}) hold and  $$ \w
\pi(t,\o)^\top =\nu(t,\o)\t(t,\o)^\top\s(t)^{-1},
\quad\hbox{where}\quad
\nu(t)=\frac{|\xi(t,\o)\s(t,\o)^\top|}{|\t(t,\o)|},
$$
if $\t(t,\o)\neq 0$, where $\xi(t,\o)=\xi(t,\o)$ is a random
$n$-dimensional $\F_t$-adapted process such that $|\xi(t,\o)|\le
\sup_{t,\o}|\pi(t,\o)|$. The constant  $C>0$ in
 (\ref{MFTineq2}) depends only on $n$ and $\s(\cdot)$.
\end{lemma}
\par
\begin{corollary}\label{k-optimal1}  Lemma \ref{k-optimal} holds if
the conditions on $\mu$ are replaced by the following condition:
 there exists $\e>0$ such that  $\mu(t)=(r(t),\ww a(t),\s(t))$ is predictable with time
 horizon $\e$, meaning that $\mu(t+\tau)$ is $\F_t$-measurable for any
 $\tau\le\e$. Then  Lemma \ref{k-optimal} holds, i.e, the Mutual Fund Theorem
holds in the sense of Lemma \ref{k-optimal}.
\end{corollary}
\par
{\it Proof of Lemma \ref{k-optimal}.} Let us continue the proof of
Lemma \ref{k-optimal}. It suffices to prove that, for any $\d>0$ and
strategy $\pi\in\Sigma(0,T)$ there exists a strategy
$\w\pi\in\Sigma_{MFT}(0,T)$ such that (\ref{MFTineq}) holds.
\par
Clearly, it suffices to prove that, for all $z\in (0,+\infty)$, for
any $\d>0$, any $m\in\{0,1,...,N-1\}$, and any
$\pi=\pi(z)\in\Sigma_\e(t_m,T)$, there exists $\w\pi=\w\pi(z)\in
\Sigma_{\e, MFT}(t_m,T)$ such that \baa \E\{ U(\ww
X(T,t_{m},z,\pi))|\F_{t_{m}}\}\le\E\{ U(\ww
X(T,t_{m},z,\w\pi))|\F_{t_{m}}\}+\frac{N-m}{N}\d. \label{ind}\eaa
 We will use mathematical induction with decreasing $m$.
First,   the statement of lemma holds for $m=N-1$ by Lemma
\ref{lemmaDet} applied on the conditional probability space. It
suffices to prove that if the statement of Lemma holds for some
$m+1\le N$ then it implies that the statement of lemma holds for
$m$.
\par
Let $z\in(0,+\infty)$ be given, and  $\pi=\pi(z)\in\Sigma_\e(t_m,T)$
be a strategy.
\par
Let $V_N(x)=U(x)$. For $x\in\R$, for $k=N-1,N-2,...$, consider a
sequence of functions $\w\pi_k:\R\times[t_k,t_{k+1}]\times\O\to\R^n$
and $V_k(x,\o):\R\times\O\to\R$ such that $\w\pi_k(x,\cdot)\in
\Sigma_{MFT}(t_k,t_{k+1})$ for any $x$ and such that \baaa &&\E\{
V_{k+1}(\ww
X(t_{k+1},t_{k},x,\w\pi_k(x,\cdot)))|\F_{t_k}\}\\&&\hspace{3.5cm}\ge\E\{
V_{k+1}(\ww X(t_{k+1},t_k,x,\pi))|\F_{t_k}, \ww
X(t_k,t_m,z,\pi)=x\}-\frac{\d}{N} \quad\hbox{a.s},\\
 &&V_k(x)\defi \E\{U(\ww X(T,t_k,x,\ww\pi_k(x,\cdot))|\F_{t_k}\},
\eaaa where $\ww\pi_k(x,\cdot)\in \Sigma_{MFT}(t_k,T)$ is  such that
\baaa \ww\pi_{k+l}(x,t)=\w\pi_{k+l}(\w
X(t_{k+l},t_k,x,\ww\pi),t),\quad \sup_{x,t,\o}|\s(t,\o)^\top
\w\pi(x,t,\o)|\le \sup_{x,t,\o,\xi}|\s(t,\o)^\top \xi|,\quad\\ t\in
[t_{k+l},t_{k+l+1}],\quad  l=0,1,..,N-k-1.
 \label{MFTineq}
 \eaaa
Here supremums are taken over $x>0$, $t\in[t_k,t_{k+1}]$, $\o\in\O$,
and over $\xi\in\R^n$ such that $|\xi|\le \sup_{t,\o}|\pi(t,\o)|$.
\par
 These functions can be constructed recursively for
 $k=N-1,N-2,...,m$.
 \par
 Existence of $\pi_k$ for every steps follows from Lemma \ref{lemmaDet}
 applied on the corresponding conditional probability space.
\par Consider the strategy \baaa
  &&\w\pi=\w\pi(z,\cdot)\quad
 \hbox{such that $\w\pi(t)=\ww \pi_k(x,t)=\w\pi_k(\w X(t_k,t_m,x,\w\pi),t)$ for $[t_k,t_{k+1}]$.}\eaaa
\par
 Let $\Pi(t)=\Pi(t,t_m,z)\defi \pi(t)\ww X(t,t_m,z,\pi)$.
We have that, for any strategy $\pi$, \baaa \ww X(T,t_m,z,\pi)&=&z+
\int_{t_m}^{T}\Pi(t)^\top\ww{\bf S}(t)^{-1}d\ww S(t)\\&=&z+
\int_{t_m}^{T}\Pi(t)^\top (\ww a(t)dt+\s(t)dw(t)). \eaaa Let
$\pi_m\defi \pi|_{[t_m,t_{m+1}]}$. It follows that  \baaa \ww
X(T,t_m,z,\pi)= \xi_{m+1}(\pi_m,z)+ \int_{t_{m+1}}^{T}\Pi(t)^\top
(\ww a(t)dt+\s(t)dw(t))\\=\ww X(T,t_{m+1},\xi_{m+1}(\pi_m,z),\pi),
\eaaa where
 \baaa\xi_{m+1}(\pi_m,z) \defi\ww
X(t_{m+1},t_m,z,\pi)= z+ \int_{t_m}^{t_{m+1}}\Pi(t)(\ww
a(t)dt+\s(t)dw(t)). \eaaa Further, \baaa \E\{ U(\ww
X(T,t_{m},z,\pi))|\F_{t_{m}}\}=\E\{\E\{ U(\ww
X(T,t_{m},z,\pi))|\F_{t_{m+1}}\}|\F_{t_m}\}\\= \E\{\E\{
U(\xi_{m+1}(\pi_m,z)+\int_{t_{m+1}}^{T}\Pi(t)(\ww
a(t)dt+\s(t)dw(t))) |\F_{t_{m+1}}\}|\F_{t_m}\}.
 \eaaa
 The equalities and inequalities here holds a.s., as well as
 inequalities and equalities for conditional expectations below.
 \par
 By the definitions and by the induction assumption that (\ref{ind}) holds with $m$ replaced by $m+1$,
 we obtain that
 \baaa
&&\E\{ U(\ww X(T,t_{m+1},\xi_{m+1}(\pi_m,z),\pi))
|\F_{t_{m+1}}\}\nonumber\\&&= \E\{
U(\xi_{m+1}(\pi_m,z)+\int_{t_{m+1}}^{T}\Pi(t)^\top(\ww
a(t)dt+\s(t)dw(t))) |\F_{t_{m+1}}\}\nonumber\\&&\le
V_{m+1}(\xi_{m+1}(\pi_m,z))+\frac{N-m-1}{N}\d. \eaaa Hence \baa \E\{
U(\ww X(T,t_{m},z,\pi))|\F_{t_{m}}\} \le \E\{
V_{m+1}(\xi_{m+1}(\pi_m,z))|\F_{t_m}\}+\frac{N-m-1}{N}\d.
\label{in1}
 \eaa Further,
by the
 choice of $\w\pi_m$, we obtain that
 \baa
 &&\E\{V_{m+1}(\xi_{m+1}(\pi_m,z))|\F_{t_{m}}\}=\E\{ V_{m+1}(\ww
X(t_{m+1},t_{m},z,\pi))|\F_{t_{m}}\}\nonumber\\
&&\le \E\{ V_{m+1}(\ww X(t_{m+1},t_{m},
z,\w\pi))|\F_{t_{m}}\}+\frac{\d}{N}\nonumber\\
&&= \E\{
V_{m+1}(\xi_{m+1}(\w\pi_m,z))|\F_{t_{m}}\}+\frac{\d}{N}.\hphantom{xxx}\label{in2}
\eaa By the definitions,
  \baa
V_{m+1}(\xi_{m+1}(\w\pi_m,z))&=&V_{m+1}(\ww X(t_{m+1},t_{m},
z,\w\pi),z)\nonumber\\&=&\E\{ U(\ww X(T,t_{m+1},\ww X(t_{m+1},t_{m},
z,\w\pi), \w\pi))|\F_{t_{m+1}}\}.\label{eq4} \eaa By the version of
the Markov property described in Theorem II.9.4 from Krylov (1980)
and applied on the conditional space given $\F_{t_m}$, we have that
the right hand part of equality (\ref{eq4}) can be rewritten as
\baa\E\{ V_{m+1}(\xi_{m+1}(\w\pi_m,z))|\F_{t_{m}}\}= \E\{ U(\ww
X(T,t_{m},z,\w\pi)|\F_{t_{m}}\}. \label{in3} \eaa We used here that
$\mu_\e(\cdot)$ is independent from $w(\cdot)$.  By
(\ref{in1})-(\ref{in3}), it follows that \baaa \E\{ U(\ww
X(T,t_{m},z,\pi))|\F_{t_{m}}\}\le\E\{ U(\ww
X(T,t_{m},z,\w\pi))|\F_{t_{m}}\}-\frac{N-m}{N}\d. \eaaa Since it
holds for any $\pi\in\Sigma(t_m,T)$, it follows that Lemma
\ref{k-optimal} holds. $\Box$
\begin{lemma}\label{lemmabounded} Theorem
\ref{MFT} holds under additional condition that
$\sup_{x>0}U_k(x)<+\infty$  in (\ref{U}) for $k=0$ and $k=N+1$.
\end{lemma}
\par
{\it Proof.} Let $t\land s=\min(t,s)$,\baaa &&r_{\e}(t)\defi
\frac{1}{\e}\int_{(t-2\e)\land 0}^{(t-\e)\land 0}r(s)ds, \quad
a_{\e}(t)\defi \frac{1}{\e}\int_{(t-2\e)\land 0}^{(t-\e)\land
0}a(s)ds, \quad \s_{\e}(t)\defi \frac{1}{\e}\int_{(t-2\e)\land
0}^{(t-\e)\land 0}\s(s)ds,\eaaa and let \baaa &&\mu_{\e}(t)\defi
(r_{\e}(t),\ww a_{\e}(t),\s_{\e}(t)),\quad  \ww a_{\e}(t)\defi
a_{\e}(t)-r_{\e}(t),\quad \t_{\e}(t)\defi\s_\e(t)^{-1}\ww a_{\e}(t).
\eaaa
\par
Consider a sequence $\e=\e_N=1/N\to 0$, $N=1,2,...$. For every
$\e=\e_i$, consider a finite sequences of times $\{t_j\}_{j=0}^{N}$
such that $t_{k+1}=t_k+\e$.
\par
 Let $\F^{\mu,\e}_t$ be the filtration
generated by $\mu_{\e}(t)$ and let $\F^\e_t$ be the filtration
generated by $(\mu_{\e}(t),w(t))$.
\par
 Let $\ww\Sigma(0,T)$ be the class of
all $\F^\e_t$-adapted processes
  $\pi(\cdot)=(\pi_1(\cdot),\ldots,\pi_n(\cdot)):[0,T]\times \O\to\R^n$ such that
   $\sup_{t,\o}|\pi(t,\o)|<+\infty$ and that if $\t_\e(t)=0$ then $\pi(t)=0$.
\par
Further, let $\Sigma_{\e,MFT} (0,T)$ denote the set of strategies
from $\Sigma_\e(0,T)$ that have the form
 $\pi(t)=\nu(t)\s_\e (t)^{-1}\t_\e(t)$, where $\nu_{\e}(t)$ is an one
 dimensional process adapted to $\F^\e_t$.
\par
For $\e>0$, let  \baaa J_\e(\pi)\defi \E U(\ww
X_\e(T,0,X_0,\pi)),\eaaa where $\ww X_\e(T,0,X_0,\pi)$ is the
discounted wealth for the model with $\mu$ replaced by $\mu=\mu_\e$
for the strategy $\pi$  given that $\ww X(0)=X_0$. The case of
$\e=0$ corresponds to the original model; in this case, the
discounted wealth is denoted as $\ww X(T,0,X_0,\pi)$.
\par
Note that the market models with $\mu=\mu_\e$ are such that
assumptions of Lemma \ref{k-optimal} are satisfied for $\e>0$.
\par
 Let $\d>0$ be given. Let $\pi\in \Sigma(0,T)$ be such that
 \baaa J(\pi)\ge
\inf_{\pi\in\Sigma(0,T)}J(\pi)-\frac{\d}{4}. \eaaa  Let $\ww X(t)=
\ww X(T,0,X_0,\pi).$  By the choice of $\Sigma(0,T)$, we have that
$C_\pi\defi\sup_{t,\o}|\pi(t,\o)|<+\infty$. Let $$\pi_{\e}(t)\defi
\frac{1}{\e}\int_{(t-2\e)\land 0}^{(t-\e)\land 0}\pi(s)ds.$$
\par
Clearly, $\pi_\e\in\Sigma_\e(0,T)$. By Lemma 3 from Shilov and
Gurevich (1967), Chapter IV, Section 5, it follows that \baaa
\mu_{\e}\to \mu,\quad \pi_{\e}\to \pi\quad \hbox{as}\quad \e\to
0+\quad\hbox{a.e. on}\quad [0,T]\times \O. \eaaa We have that \baa
&&\ww X(T,0,X_0,\pi)=X_0+\int_{0}^{T}\ww
X(t,0,X_0,\pi)\pi(t)^\top\ww{\bf S}(t)^{-1}d\ww S(t) \nonumber\\&&=
X_0\exp\left[\int_{0}^{T} \pi(t)^\top\ww
a(t)dt-\frac{1}{2}\int_{0}^{T}|\pi(t)^\top\s(t)|^2dt+\int_{0}^{T}\pi(t)^\top\s(t)
dw(t)\right].\hphantom{xxxxx}\label{X0} \eaa Similarly, \baa &&\ww
X_{\e}(T,0,X_0,\pi_\e)\nonumber\\&&= X_0\exp\left[\int_{0}^{T}
\pi_\e(t)^\top\ww
a_\e(t)dt-\frac{1}{2}\int_{0}^{T}|\pi_\e(t)^\top\s_\e(t)|^2dt+\int_{0}^{T}\pi_\e(t)^\top\s_\e(t)
dw(t)\right].\hphantom{xxxxx} \label{Xe}\eaa
\par
Let $Y_{\e,\e}(t)\defi \log X_\e(t,0,X_0,\pi_\e)$ and $Y(t)\defi
\log X_\e(t,0,X_0,\pi_\e)$.
\par
 Clearly, \baa\E|Y_{\e,\e}(T)- Y(T)|^2\to 0\quad\hbox{as}\quad
\e\to 0.\label{elog}\eaa It follows that there exists a subsequence
$\{\e_i\}$ such that \baa Y_{\e,\e}(T)\to Y(T)\quad \hbox{a.s.}\quad
\hbox{as}\quad \e=\e_i\to 0.\label{subs}\eaa By the assumptions, all
functions $U_k$ are bounded. By Lebesgue's Dominated Convergence
Theorem, this subsequence $\{\e_i\}$ is such that \baa \E|U_k(\ww
X_\e(T,0,X_0,\pi))-U_k(\ww X(T,0,X_0,\pi)|^2\to 0\quad\hbox{as}\quad
\e=\e_i\to 0\nonumber\\ k=0,1,...,N+1.\label{eU}\eaa
\par
By (\ref{eU}), it follows that \baa \E U_0(\ww X_\e(T,0,X_0,\pi))\to
\E U_0(\ww X(T,0,X_0,\pi)\quad\hbox{as}\quad \e=\e_i\to
0.\hphantom{xxx}\label{eU10}\eaa
\par
By (\ref{eU}) and (\ref{elog}), it follows that \baa \E U_{N+1}(\ww
X_\e(T,0,X_0,\pi_\e))Y_{\e,\e}(T)\to \E U_{N+1}(\ww
X(T,0,X_0,\pi)Y(T)\quad\hbox{as}\quad \e=\e_i\to
0.\hphantom{xxx}\label{eU1}\eaa Further, let $k\in{1,...,N}$, and
let \baaa z_{\e,\e}(t)\defi X_\e(t,0,X_0,\pi_\e)^{\d_k}= \exp(\d_k
Y_{\e,e}(t)),\quad z(t)\defi X(t,0,X_0,\pi)^{\d_k}=\exp(\d_k
Y(t)).\eaaa  By Ito formula, we obtain \baaa
dz_{\e,\e}(t)=z_{\e,\e}(t)\left(\d_kdY_{\e,\e}(t)+\frac{1}{2}\d_k^2|\pi_{\e,\e}(t)^\top\s_\e(t)|^2dt\right),
\quad z_{\e,\e}(0)=X(0)^{\d_k}, \eaaa where $$ dY_\e(t)=
\pi_\e(t)^\top\ww
a_\e(t)dt-\frac{1}{2}|\pi_\e(t)^\top\s_\e(t)|^2dt+\pi_\e(t)^\top\s_\e(t)
dw(t). $$
 Similarly, we obtain \baaa
dz(t)=z(t)\left(\d_kdY(t)+\frac{1}{2}\d_k^2|\pi(t)^\top\s(t)|^2dt\right),
\quad z(0)=X(0)^{\d_k}, \eaaa
 where $$ dY(t)=
\pi(t)^\top\ww
a(t)dt-\frac{1}{2}|\pi(t)^\top\s(t)|^2dt+\pi(t)^\top\s(t) dw(t).
$$
By Theorem II.8.1 from Krylov (1980), p.102, we have that
$\E|z_{\e,\e}(T)-z(T)|^2\to 0$ as $\e=\e_i\to 0$ for any
$k=1,...,N$. By (\ref{eU}), we obtain for $k=1,...,N$ that \baa\E
U_{k}(\ww X_\e(T,0,X_0,\pi_\e))z_{\e,\e}(T)\to \E U_{k}(\ww
X(T,0,X_0,\pi)z(T)\quad\hbox{as}\quad \e=\e_i\to
0.\hphantom{xxx}\label{eU3}\eaa
\par
By (\ref{eU10})-(\ref{eU3}), we obtain that \baa J_\e(\pi_\e)=\E
U(\ww X_\e(T,0,X_0,\pi_\e))\to J(\pi)=\E U(\ww
X(T,0,X_0,\pi)\quad\hbox{as}\quad \e=\e_i\to 0.\hphantom{xxx}
\label{eU20}\eaa
 It follows
that there exists $N_1>0$ such that, for every $i\ge N_1$, \baaa
J_\e(\pi_\e)\ge J(\pi)-\frac{\d}{4},\quad \e=\e_i. \eaaa
 \par
  Let
$\w\pi_{\e,\e}\in \Sigma_{\e,MFT}(0,T)$ be the strategy defined in
Lemma \ref{k-optimal} as a strategy that outperform  the strategy
$\pi_\e$ for the market with $\mu=\mu_\e$, i.e., such that
$\nu_{\e}(t)$ is $\F^{\e}_t$-adapted process and \baaa
J_\e(\pi_{\e,\e})\ge J_\e(\pi_\e)-\frac{\d}{4}. \eaaa Following the
proof of Lemma \ref{lemmaDet} we obtain similarly to (\ref{fopt})
that, if $\t(t)\neq 0$, then \baa \pi_{\e,\e}(t)^\top
=\t_\e(t)^{\top}\s_\e(t)^{-1}\nu_\e(x,t), \quad \hbox{where}\quad
\nu_\e(t)= \frac{|\xi_\e(t,\o)^\top\s_\e(t)|}{|\t_\e(t)|},
 \label{fopte}\eaa
  and
 where $\xi_\e(t,\o)$ is a $n$-dimensional vector such that  $|\xi_\e(t,\o)|\le |\pi_\e(t,\o)|$.
If $\t(t)= 0$ then $\w\pi_{\e,\e}(t)=0$.
\par By estimate (\ref{MFTineq2}) in Lemma \ref{k-optimal}, we have
that \baa &&\sup_{t,\o,\e}|\pi_{\e,\e}(t,\o)| \le
C\sup_{t,\o,\e}|\pi_\e(t,\o)|\le
C\sup_{t,\o}|\pi(t,\o)|,\label{inn2} \eaa where  $C=C(n,\s)>0$ is a
constant.
\par
Let \baaa &&\pi_{\e,0}(t)^\top\defi\frac{|\t_\e(t)|}{|\t(t)|}
\nu_{\e}(t)\t(t)^\top\s(t)^{-1}
\quad \hbox{if}\quad \t(t)\neq 0,\ \t_\e(t)\neq 0,\\
&&\pi_{\e,0}(t)=0 \quad \hbox{if}\quad \t(t)= 0\quad\hbox{or}\quad
\t_\e(t)= 0. \eaaa It follows that, if $\t(t)\neq 0$, $\t_\e(t)\neq
0$\baaa \pi_{\e,0}(t)^\top=\frac{|\t_\e(t)|}{|\t(t)|}\,
\frac{|\xi_\e(t,\o)^\top\s_\e(t)|}{|\t_\e(t)|} \t(t)^\top\s(t)^{-1}=
\frac{|\xi_\e(t,\o)^\top\s_\e(t)|}{|\t(t)|} \t(t)^\top\s(t)^{-1}.
\eaaa  Hence \baa \sup_{t,\o,\e}|\pi_{\e,0}(t,\o)|\le
C\sup_{t,\o,\e}|\pi_\e(t,\o)|\le C\sup_{t,\o}|\pi(t,\o)|,
\label{inn3}\eaa where $C=C(n,\s)$ is a constant that depends only
on $n$ and $\s$.
\par
 The equations for  $\ww X_\e(T,0,X_0,\pi_{\e,\e})$ and $\ww X_\e(T,0,X_0,\pi_{\e,0})
$ are similar to equations (\ref{X0})-(\ref{Xe}).  Clearly,
$\pi_{\e,\e}(t,\o)-\pi_{\e,0}(t,\o)\to 0$ a.e.. Using
(\ref{inn2})-(\ref{inn3}), we obtain that $\E| \log\ww
X_\e(T,0,X_0,\pi_{\e,\e})- \log \ww X(T,0,X_0,\pi_{\e,0})|^2\to 0$
as $\e\to 0$. It follows that there exists another  subsequence
$\{\e_i\}$   (a subsequence of the subsequence from (\ref{subs}))
such that $\e_i\to 0$ and $\log\ww X_\e(T,0,X_0,\pi_{\e,\e})-\log
\ww X(T,0,X_0,\pi_{\e,0})\to 0$ a.s. as $\e=\e_i\to 0$.  Similarly
to (\ref{eU10})-(\ref{eU20}), we obtain that this subsequence
$\{\e_i\}$ is such that \baaa J_\e(\pi_{\e,\e})-J_\e(\pi_{\e,0})=\E
U(\ww X_\e(T,0,X_0,\pi_{\e,\e}))-\E U(\ww X(T,0,X_0,\pi_{\e,0})\to
0\eaaa as $\e=\e_i\to 0$. It follows that there exists $N>N_1>0$
such that, for every $i\ge N$, \baaa J_0(\pi_{\e,0})\ge
J_\e(\pi_{\e,\e})-\frac{\d}{4},\quad \e=\e_i. \eaaa Finally, we
obtain that \baaa J_0(\pi_{\e,0})\ge
J_\e(\pi_{\e,\e})-\frac{\d}{4}\ge J_\e(\pi_{\e})-\frac{\d}{2}\ge
J_0(\pi)-\frac{3\d}{4},\quad \e=\e_i. \eaaa This completes the proof
of Lemma \ref{lemmabounded}.
 $\Box$
\par
\def\J{{\cal J}}
{\it Proof of Theorem \ref{MFT}.}
 It suffices to show that there exists $C>0$ such that, for any $\d>0$ and $\pi\in\Sigma(0,T)$, there exists
$\w\pi\in\Sigma_{MFT}(0,T)$ such that
(\ref{MFTineq0})-(\ref{MFTineq2}) hold.
\par
For $K>0$, let $U^{(K)}(x)$ be defined by (\ref{U}) with $U_0$
replaced by $\min(U_0(x),K)$ and with $U_{N+1}$ replaced by
$\min(U_{N+1}(x),K)$. Let  $\J_K(\pi)=\E U^{(K)}(\ww
X(T,0,X_0,\pi)$.
 \par Let $C>0$ be the constant (\ref{MFTineq2}) that exists by Lemma
 \ref{lemmabounded} for all $K>0$. (Note that this constant does not depend on $K$). Let
$\d>0$ and $\pi\in\Sigma(0,T)$ be given, Clearly, there exists $K>0$
such that $\J_K(\pi)\ge J(\pi)-\d/2$. By Lemma \ref{lemmabounded},
there exists $\w\pi\in \Sigma_{MFT}(0,T)$ such that  $\J_K(\w\pi)\ge
\J_K(\pi)-\d/2$ and (\ref{MFTineq2}) holds. In addition, we have
that $\J_K(\w\pi)\ge J(\w\pi)$ for large enough $K$ (it suffices to
take $K>\sup_{x\in(0,1)}U_{N+1}(x)$). For these $K$, we have that
$J(\w\pi)\ge \J_K(\pi)\ge J(\pi)-\d$. Then the proof follows. $\Box$
\section{Discussion and comments}
\begin{enumerate}
\item
Theorem \ref{MFT}  represents a weakened version of Mutual Fund
Theorem since  it states only suboptimality of the strategies from
the required class. A stronger version of this theorem is known for
many special cases. In particular, there are stronger versions of
Lemma \ref{lemmaDet}; see, e.g., Khanna and Kulldorff (1999),
Dokuchaev and Haussmann (2001), Schachermayer {\it et al} (2009).
%\begin{itemize}\item
Let us explain why  these versions of Lemma \ref{lemmaDet} cannot be
applied in our proof.
\par
Khanna and Kulldorff (1999) proved that a strategy from a class
similar to $\Sigma_{MFT}$  can outperform any Markov strategies. Our
setting with random parameters requires to include strategies that
are not necessary Markov.
%\item
\par
 Schachermayer {\it et al} (2009)) found that the Mutual Fund
Theorem holds for a market where claims $F(Z(T))$ can be replicated
via trading of a mutual fund with the discounted price $Z(t)$ for
deterministic functions $F$. Here  $Z(t)$ is the log-optimal
discounted wealth such that \baaa dZ(t)=Z(t)\t(t)^\top
\s(t)^{-1}{\bf S}(t)^{-1}dS(t), \quad Z(0)=1. \eaaa In the same
framework,
 Dokuchaev and
Haussmann (2001) found that Mutual Fund Theorem holds in a more
special case, when the scalar value $\int_0^T|\t(t)|^2dt$ is
non-random. In this case, there is the required replicability of
claims $F(Z(T))$. However,  these results cannot replace Lemma
\ref{lemmaDet}, because they require certain special properties for
$U$ and for the functions $V_m$ in the proof of Lemma
\ref{k-optimal}. If we assume  these properties for $U$, it is not
clear how to prove that they will be transferred to $V_m$.
%\end{itemize}
\item
It can be seen from the proofs  above that, in many cases of random
$\mu$, the suboptimal terminal discounted wealth cannot be presented
as $F(Z(T))$ for a deterministic function $F:\R\to\R$. Respectively,
these cases are not be covered by the method based on the
replication of these claims (Schachermayer {\it et al} (2009),
Dokuchaev and Haussmann (2001)).
\item
The condition (\ref{U}) in Theorem \ref{MFT} restricts the choice of
singularity for admissible utility functions $U$ at $x=0$. However,
this condition is rather technical; we need it ensure the transfer
from the market model from Lemma \ref{k-optimal} to the more general
market model in Theorem \ref{MFT}. However, the model in Lemma
\ref{k-optimal} is quite reasonable itself, since it is natural to
assume some stability and predictability of the parameters of the
distributions; this assumption is required by any statistical
analysis. There are many well developed methods that may help to
forecast the market parameters on a small enough horizon $\e>0$; in
particular, a frequency criterion of predictability on a finite
horizon can be found in Dokuchaev (2010).
\item
It can be noted also that the construction of  suboptimal strategies
from the proof above  shows  that,  in the general case, these
strategies cannot be presented as $\pi(t)=f(X(t),S(t),\mu(t),t)$,
where  $f$ is a deterministic function. This means that dynamic
programming method cannot be applied directly to this model.
\item
In our setting, we assumed that the admissible strategies are such
that if $\t(t)=0$ then $\pi(t)=0$.  In fact, our version of Mutual
Fund Theorem does not necessary hold for a wide class without this
restriction given our class of utilities. For instance, for a convex
function $U(x)=x^2$ and $\t(t)\equiv 0$, the only strategy
$\pi_{MFT}$ from Mutual Fund Theorem is zero; however, this strategy
is outperformed by any non-trivial strategy.
\end{enumerate}
\section*{References}
$\hphantom{xx}$ Brennan, M.J. (1998). { The role of learning in
dynamic portfolio decisions.} {\em European Finance Review} {\bf 1},
295--306.
\par
Dokuchaev, N., and Haussmann, U.  (2001). Optimal portfolio
selection and compression in an incomplete market.  {\em
Quantitative Finance}, {\bf 1}, 336-345.

Dokuchaev, N. (2010). Mean variance and goal achieving portfolio for
discrete-time market with currently observable source of
correlations. {\it ESAIM: Control, Optimisation and Calculus of
Variations}, in press.

 Dokuchaev, N. (2010). Predictability on finite horizon
for processes with exponential decrease of energy on higher
frequencies.
 {\it Signal processing} {\bf 90} Issue 2, February 2010, 696--701 (in press).

Feldman, D. (2007). Incomplete Information Equilibria: Separation
Theorems and Other Myths. {\it Annals of Operations Research} {\bf
151}, 119--149.

Karatzas, I., and Shreve, S.E. (1998).   {\em Methods of
Mathematical Finance}, Springer-Verlag, New York.

Khanna A.,, Kulldorff M. (1999). A generalization of the mutual fund
theorem. {\it Finance and Stochastics} {\bf  3}, 167–185 (1999).

Krylov, N.V.  (1980).  {\em Controlled diffusion processes}.
Shpringer-Verlag.

Li, D.,  and Ng, W.L. (2000). Optimal portfolio selection:
multi-period mean--variance optimization. {\em Mathematical Finance}
{\bf 10} (3),  387--406.

Lim, A. (2004). Quadratic hedging and mean-variance portfolio
selection with random parameters in an incomplete market. {\it
Mathematics of Operations Research} {\bf 29}, iss.1, 132 - 161.

Lim, A., and Zhou, X.Y. (2002).
 Mean-variance portfolio selection with random
parameters in a complete market. {\it Mathematics of Operations
Research} {\bf 27}, iss.1, 101 - 120.

 Schachermayer, W., Sîrbu, M., Taflin, E. (2009). In
which financial markets do mutual fund theorems hold true? {\it
 Finance and Stochastics} {\bf 13},
49--77.

Shilov G.E, Gurevich B.L (1967). {\it Integral, measure and
derivative: a unified approach.} Nauka, Moscow.

\end{document}